\documentclass{llncs}
\usepackage{graphicx}
\usepackage{amsmath,amssymb} 
\usepackage{color}
\usepackage[width=135mm,left=23mm,paperwidth=176mm,height=210mm,top=18mm,paperheight=246mm]{geometry}

\usepackage{acronym}

\usepackage{hyperref}

\usepackage{minted}

\usepackage{xcolor}

\usepackage{tabularx}
\usepackage{booktabs}

\acrodef{DDD}{Domain-Driven Design}
\acrodef{DM}{Domain Model}
\acrodef{DSL}{Domain Specific Language}
\acrodef{LLMs}{Large Language Models}
\acrodef{LoRA}{Low Rank Adaptation}
\acrodef{PEFT}{Parameter Efficient Fine-Tuning}
\acrodef{QLoRA}{Quantized Low Rank Adaptation}
\acrodef{SME}{Small and Medium-Sized Enterprises}
\acrodef{UML}{Unified Modeling Language}

\begin{document}

\pagestyle{headings}

\mainmatter

\title{Leveraging Generative AI for Enhancing Domain-Driven Software Design}

\author{Götz-Henrik Wiegand\inst{1} \and Filip Stepniak\inst{2} \and Patrick Baier\inst{1}}

\institute{Hochschule Karlsruhe University of Applied Sciences, Germany\\ 
\email{goetz.henrik.wiegand@gmail.com \& patrick.baier@h-ka.de}
\and
esentri AG, Ettlingen, Germany\\
\email{filip.stepniak@esentri.com}
}

\maketitle

\begin{abstract}

Domain-Driven Design (DDD) is a key framework for developing cus\-to\-mer-oriented software, focusing on the precise modeling of an application's domain. Traditionally, metamodels that describe these domains are created manually by system designers, forming the basis for iterative software development. This paper explores the partial automation of metamodel generation using generative AI, particularly for producing domain-specific JSON objects. By training a model on real-world DDD project data, we demonstrate that generative AI can produce syntactically correct JSON objects based on simple prompts, offering significant potential for streamlining the design process. To address resource constraints, the AI model was fine-tuned on a consumer-grade GPU using a 4-bit quantized version of Code Llama and Low-Rank Adaptation (LoRA). Despite limited hardware, the model achieved high performance, generating accurate JSON objects with minimal post-processing. This research illustrates the viability of incorporating generative AI into the DDD process, improving efficiency and reducing resource requirements, while also laying the groundwork for further advancements in AI-driven software development.

\end{abstract}



\section{Introduction} 



Creating customer-oriented software demands efficient tools and methods. A promising approach is the \ac{DDD} pattern\cite{evansDomaindrivenDesignTackling2004}, a robust framework for software development emphasizing the understanding and modeling of the application's domain. Initially, the software is described using \ac{DSL} in JSON or UML, forming a \ac{DM} that underpins the iterative development process. From this \ac{DM}, a code framework is derived, which is then endowed with logic to create a prototype. This prototype generates insights for refining the \ac{DM} further.  

The initial \ac{DM} generation is typically a manual task performed by a system designer using a GUI tool. To enhance this process, we explore in this paper the possibility of partially automating it with the help of generative AI. We demonstrate how generative models can learn to create syntactically correct JSON objects for describing the \ac{DDD} \ac{DM}. Moreover, we show that being trained on real-world data from existing \ac{DDD} projects, the AI model can automatically generate new parts of a \ac{DM} through simple interactions with a system prompt.
The AI model's ability to produce syntactically correct JSON objects ensures machine readability, facilitating integration into existing \ac{DDD} development tools. 


Due to data confidentiality, the use of commercial \ac{LLMs} are not an option, which led us to the constraint to develop the JSON code generator model on resource-restrictive hardware, specifically a single consumer-grade GPU. 

The final results on the test dataset yielded impressively low loss on JSON generation and high BLEU\cite{papineni-etal-2002-bleu} scores, underscoring the model's proficiency. Most of the generated JSON objects exhibited syntactical correctness with minimal post-processing, and all JSON objects created from clear prompts were syntactically correct. The successful creation of a code generator for JSON objects in the \ac{DSL} signifies a pivotal advancement towards incorporating generative AI into the \ac{DDD}-based software development process, enhancing both efficiency and efficacy.

\section{Related Work} 

The foundation for this work lies in the principles of \ac{DDD}, as established by Eric Evans in his seminal works\cite{evansDomaindrivenDesignTackling2004,evansDomainDrivenDesignReference2014}. \ac{DDD} provides a strategic approach to software development, emphasizing the modeling of complex systems based on their underlying business domains. The company internal framework used for this work builds upon these principles and knowledge regarding this were obtained from the internal documentation\cite{esentriInternalCompanyDocumentation}.

To address the challenges of efficient resource utilization in AI model training and deployment, techniques such as \ac{PEFT} were employed, specifically the \ac{LoRA} method introduced by Hu et al.\cite{huLoRALowRankAdaptation2021}. Further refinements, including quantization methods like \ac{QLoRA} by Dettmers et al.\cite{dettmersQLoRAEfficientFinetuning2023}, played a crucial role in optimizing performance on resource-constrained hardware. The model in this work was quantized to 4-bit precision using the `BitsAndBytes` library from Hugging Face\cite{HuggingFaceBitsAndBytes}. This approach is supported by research on low-precision quantization, such as the work of Sun et al.\cite{sunUltraLowPrecision4bit} and Neshaei et al.\cite{neshaeiImpactQuantizationRobustness2024}.

The model used for the code generation component was \textit{Code Llama} from \textit{Meta}\footnote{https://www.llama.com/code-llama/}, proposed by Rozière et al.\cite{roziereCodeLlamaOpen2024}. Other models relevant to this field include StarCoder\cite{liStarCoderMaySource2023} and CodeT5\cite{wangCodeT5IdentifierawareUnified2021}. Additionally, commercial AI code generation tools such as \textit{GitHub Copilot}\cite{GitHubCopilotYour2024} and \textit{Amazon CodeWhisperer}\cite{KICodegeneratorAmazonCodeWhisperer} provide further context and reference in evaluating the landscape of AI-assisted software development.

For the evaluation of the importance of the hyperparameters after hyperparameter Tuning we used a permutation importance analysis referencing the \textit{Random Forest Regressor} from Louppe \cite{louppeUnderstandingRandomForests2015}. 

For performance evaluation, two key metrics, \textit{BLEU}\cite{tranDoesBLEUScore2019} and Loss, were used to assess the quality of the fine-tuned model. These metrics have been adopted in the evaluation of code generation models, as discussed in the works of Chen et al.\cite{chenEvaluatingLargeLanguage2021} and Yetiştiren et al.\cite{yetistirenEvaluatingCodeQuality2023}, providing a foundation for assessing syntactic and semantic alignment in generated outputs.

\section{Methods} 
In this section, a comprehensive outline of the methodological approach is presented, detailing the processes and techniques used for data handling, model development, and evaluation.

\subsubsection{Goals and Constraints:}
\label{subSec:Goals}

This work explores integrating Generative AI into the software development process within a \ac{DDD} framework, focusing on automating early-stage development by generating \ac{DM}s from business requirements. The prototype uses causal language modeling to produce \ac{UML} representations in JSON format, aligning with the iterative nature of \ac{DDD}. 

Technical constraints include the use of open-weight models due to data privacy regulations, prohibiting commercial AI models and requiring local hosting and fine-tuning. Additionally, the project operates under a €1000 budget for external computational resources, demanding resource-efficient model selection and training. The limited dataset further challenges the generation of unbiased, generalizable results, requiring mitigation of data-induced biases. The research aims to evaluate the feasibility of AI-driven code generation within these constraints, emphasizing model performance, resource management, and compliance.

\subsubsection{Data Basis:}
The dataset utilized for this study comprises 1,022 files, each containing a single JSON object. Of these, 821 files—accounting for 80\% of the dataset—originate from a customer project, while the remaining 20\% are derived from a test project. The data represent hierarchically structured \ac{DDD} logic, encoded in JSON format.

Each JSON object consists of specific key-value pairs, which are defined within a specialized framework. These key-value structures are inherited from a metamodel, which serves as the basis for the framework's logic. However, the metamodel itself is not included within the dataset, limiting direct access to the underlying inheritance structure.

\subsubsection{Data Pre-Processing:}
\label{subSec:DataPrePro}

The data pre-processing step is essential for developing a robust code generator, particularly given the dataset's significant bias, with approximately 80\% of the dataset sourced from a customer project and 20\% from a test project. This dataset, comprising 1,022 files containing completed JSON\cite{bourhisJSONDataModel2017} objects, necessitates careful handling to ensure effective model training. 

\begin{figure}[ht]
	\centering
	\fbox{\includegraphics[width=0.97\linewidth]{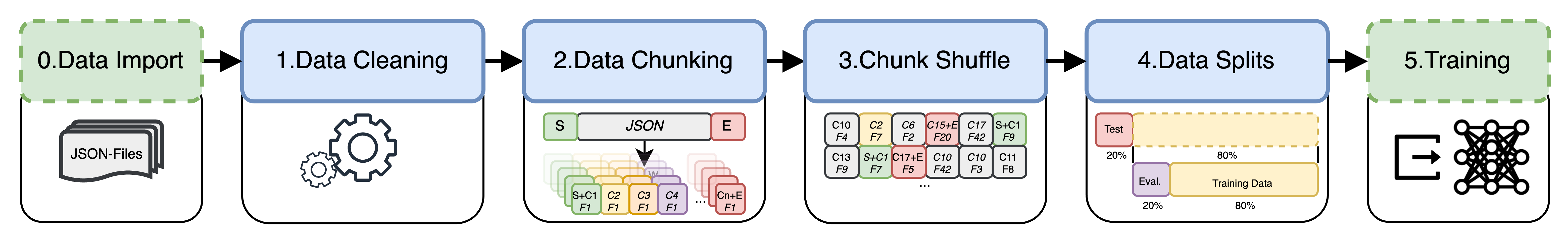}}
	\caption{Abstracted visualization of the various steps of data pre-processing with data cleaning, chunking and splitting to the various data sets for training.}
	\label{img:preprocessingflow_simple}
\end{figure}

The pre-processing process began with data import, followed by cleaning and abstraction, where high-variability keys were replaced with placeholder values to anonymize customer-specific information. This step not only protects sensitive data but also simplifies the dataset's complexity, allowing for a clearer focus on the JSON structure. Subsequently, the data was chunked into non-overlapping segments of 2,048 tokens, which were shuffled to enhance randomness. The final step involved a double 80:20 split\cite{josephOptimalRatioData2022} of the data into training, evaluation, and test sets, resulting in 64\% for training, 16\% for evaluation, and 20\% for testing. This structured approach to data pre-processing ensures that the dataset is well-prepared for effective model training while maintaining compliance with data privacy standards. After exportation, the datasets were versioned for future use, solidifying the pre-processing phase as a foundational element in the overall development process. A full process flow of data pre-processing is displayed in Figure \ref{img:preprocessingflow_simple}. 

\subsubsection{Training and Setup:}
\label{subSec:TrainingSetup}

The foundation of the code generator utilizes the Code Llama 7B model, released by \textit{Meta}\footnote{https://www.llama.com/code-llama/} proposed by Rozière et al.\cite{roziereCodeLlamaOpen2024}. With a VRAM size of approximately 25 GB, it necessitated adaptations for the limited hardware available, including a local PC with an RTX 2080 GPU (11 GB VRAM) and a \textit{Lambda Cloud}\footnote{https://lambdalabs.com/service/gpu-cloud} instance with an RTX A6000 GPU (48 GB VRAM). Due to financial constraints, the cloud instance was primarily used for hyperparameter tuning. To facilitate fine-tuning on the local setup, a 4-bit quantization was applied, reducing the model's size to around 4 GB VRAM. The LoRA method, part of \ac{PEFT}, was selected due to its proven effectiveness when combined with model quantization. The \textit{Hugging Face Transformers}\footnote{https://huggingface.co/docs/transformers/index} framework's Trainer\cite{HuggingfaceTrainer} was employed alongside a \ac{LoRA} adapter to optimize training on constrained hardware. This was used to fine-tune \textit{Code Llama 7B} to generate JSON with the help of Next Token Prediction from the dataset. Key training arguments were established to manage resource use, including batch size, gradient accumulation steps, and mixed-precision training. Evaluation metrics were critical for assessing model performance; while the built-in loss function was utilized, metrics such as \textit{BLEU}\cite{papineniBleuMethodAutomatic2002} and ROUGE-$L$-$F1$\cite{linROUGEPackageAutomatic2004} were employed to guide the training process. Memory overflow issues were addressed by implementing a custom function for pre-processing logits, ensuring efficient metric evaluation. Overall, the training utilized both local and cloud resources. 

\subsubsection{Hyperparameter Tuning:}
\label{sec:hyperparameterTuningMethods}

Hyperparameter tuning is crucial for optimizing model performance by selecting the most effective values for key hyperparameters. In this process, two categories of hyperparameters were identified for tuning: basic training parameters—learning rate, number of training epochs, and warm-up steps—and adapter-specific parameters such as the rank (R-value) and alpha value of the \ac{LoRA} adapter\cite{HuggingfaceTrainer,MethodsToolsEfficient,LoRAHuggingface}. These parameters were chosen due to their significant impact on the model's performance, particularly in hardware-constrained environments. 

To guide the tuning process, initial ranges were defined: a learning rate of 1e-5 to 5e-5, 1 to 5 training epochs, 200 to 1200 warm-up steps, an R-value of 4 to 16, and an alpha value of 4 to 16\cite{MethodsToolsEfficient,LoRAHuggingface}. These ranges were informed by engineering practices and recommendations from existing documentation. After 100 trials, adjustments were made to refine the search, particularly for the number of training epochs and the \ac{LoRA} rank. In the second phase of tuning, the R-value range was expanded to 4 to 32, and the number of training epochs increased to 5 to 12, allowing further exploration of these critical parameters while leaving the other ranges unchanged. This iterative approach helped maximize model efficiency under the available hardware constraints.


To determine the optimal hyperparameters for final training, a multi-objective weight\-ed sum approach was used, following Bazgan et al.\cite{bazganPowerWeightedSum2022}. The weighted sum function \( f(x) \) was initially defined for three evaluation metrics (Equation \eqref{weighted_sum:with_ROUGE}).

However, hyperparameter tuning results showed that the ROUGE-$L$-$F1$ metric was outside the expected range. As a result, the weight for ROUGE-$L$-$F1$ was set to zero in Equation \eqref{weighted_sum:RougeZerrroWeight}, removing its influence on \( f(x) \).

Finally, introducing the Inverse Loss \(\widetilde{L}(x)\) as \( 1 - L(x) \), the weighted sum was simplified to \( f(x) \) in Equation \eqref{weighted_sum:fx}.

\begin{align}
    f(x) &= w_{\text{Loss}} \cdot (1 - L(x)) + w_{\text{BLEU}} \cdot B(x) + w_{\text{ROUGE-L}_{\text{F1}}} \cdot R(x) \label{weighted_sum:with_ROUGE} \\
   f(x) &= w_{\text{Loss}} \cdot (1 - L(x)) + w_{\text{BLEU}} \cdot B(x) + 0 \cdot R(x) \label{weighted_sum:RougeZerrroWeight} \\
   f(x) &= w_{\text{Loss}} \cdot \widetilde{L}(x) + w_{\text{BLEU}} \cdot B(x) \label{weighted_sum:fx}
\end{align}

\subsubsection{Model Assessment}
\label{sec:ModelAssessmentMethods}

To comprehensively evaluate the model's performance for a generative \ac{DDD} system, a three-phase assessment approach was used, as traditional metrics alone offer limited insight. In the first phase, the evaluation metrics Loss and \textit{BLEU}\cite{tranDoesBLEUScore2019} from both the training and test datasets were reviewed. The second phase assessed the syntactic correctness and machine-readability of the generated JSON objects. Here, 100 JSON samples were generated from 10 \textit{clear} and 10 \textit{experimental} prompts. \textit{Clear} prompts specify a distinct DDD class object, guiding the model to create a corresponding JSON object, while \textit{experimental} prompts progressively reduce detail, giving the model more room for errors and issues. If any of the generated samples exceeded the token length limit of 4,000, post-processing was applied to ensure completeness, followed by verification through a JSON parser. The final phase involved a qualitative review of the generated JSON objects to identify potential errors and issues. This multi-step evaluation offers a more detailed understanding of the model's quality and its suitability for real-world applications.

\section{Results and Discussion} 
In this section, the results of the Hyperparameter Tuning, Final Model Training, and Model Assessment are summarized and discussed.

\subsubsection{Hyperparameter Tuning:}
Table \ref{tab:top3perEvalMetric} displays the results of the top three values for the different evaluation metrics (objectives) from the hyperparameter tuning process.

\begin{figure}[h]
	\centering
	\fbox{\includegraphics[width=0.8\linewidth]{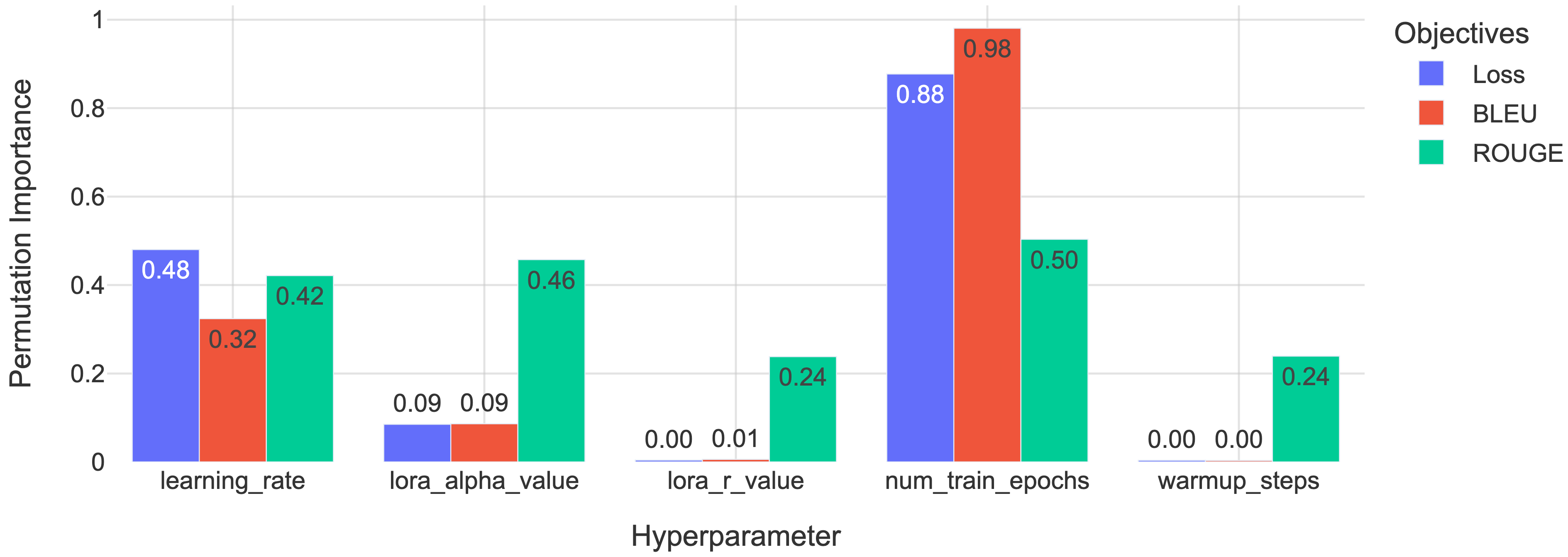}}
	\caption{Parameter importance for multiple evaluation metrics (\textit{Objectives}) with importance calculated using permutation importance with \textit{Random Forest Regressor}\cite{louppeUnderstandingRandomForests2015}.}
	\label{img:plot:ParameterImportance}
\end{figure}

The results from the hyperparameter tuning also allow for the derivation of the importance of individual hyperparameters. A Permutation Importance analysis was conducted using a \textit{Random Forest Regressor}\cite{louppeUnderstandingRandomForests2015} to assess the influence of each parameter on the evaluation metrics (objectives) shown in Figure \ref{img:plot:ParameterImportance}. This method helps to quantify how changes in specific hyperparameters affect the model's performance.

In the analysis of the results from hyperparameter tuning, a noticeable discrepancy was observed between the expected and actual values of the ROUGE-$L$-$F1$ score. The ROUGE-$L$-$F1$ score was anticipated to approach 1. During hyperparameter tuning, it reached a maximum of only approximately 0.062 in the second trial (see Table \ref{tab:top3perEvalMetric}). Due to this significant deviation, the ROUGE-$L$-$F1$ metric was excluded from the determination of the optimal hyperparameters.

\begin{figure}[h]
	\centering
	\fbox{\includegraphics[width=0.8\linewidth]{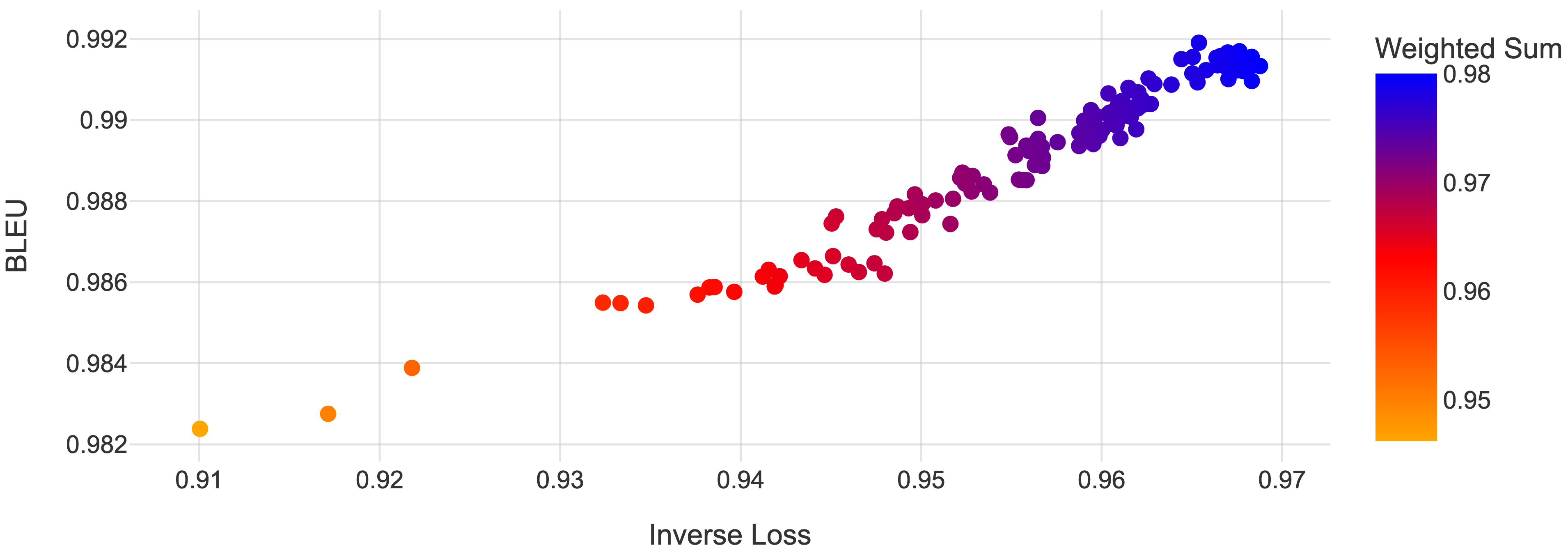}}
	\caption{Weighted sum in reference to inverse Loss $\widetilde{L}(x)$ and BLEU $B(x)$.}
	\label{img:plot:WeightedSumInReferenceToILossBLEU}
\end{figure}

Subsequently, the weighted sum method described in Section \ref{sec:hyperparameterTuningMethods} was applied and calculated for each trial. Assuming that all evaluation metrics converge towards 1 (using the inverse loss as \(1 - \text{Loss}\)), it can be inferred that the trial with the maximum weighted sum defines the optimal hyperparameters, denoted as $\theta^*$. Table \ref{tab:top5WeightedSum} presents the top five trials, ranked by their weighted sum along with their respective objectives. Figure \ref{img:plot:WeightedSumInReferenceToILossBLEU} illustrates the convergence of the weighted sum towards 1, in relation to BLEU and inverse loss, providing a visual representation of this progression.

\begin{table}[h]
    \centering
    \resizebox{\linewidth}{!}{
    \begin{tabular}{ccccccccccccc}
        \toprule
        \textbf{Trial Number} & \textbf{Ranking} & \textbf{Loss$\downarrow$} & \textbf{BLEU$\uparrow$} & \textbf{ROUGE-$L$-$F1\uparrow$} & \textbf{Learning Rate} & \textbf{LoRA Alpha} & \textbf{LoRA R} & \textbf{Train Epochs} & \textbf{Warmup Steps} \\
        \midrule
        116 & 1. Loss & \textbf{0.031224} & 0.991329 & 0.046125 & 3.4e-05 & 30 & 10 & 6 & 448 \\
        128  & 2. Loss & 0.03168 & 0.990963 & 0.04753 & 3.5e-05 & 29 & 13 & 6 & 419 \\
        127  & 3. Loss & 0.031686 & 0.991554 & 0.046529 & 4.2e-05 & 22 & 5 & 6 & 1044 \\
        \midrule
		110  & 1. BLEU & 0.034611 & \textbf{0.991905} & 0.047163 & 3.8e-05 & 17 & 10 & 12 & 1194 \\
		124  & 2. BLEU & 0.032367 & 0.991696 & 0.047422 & 3.4e-05 & 24 & 10 & 9 & 974 \\
        125  & 3. BLEU & 0.033012 & 0.991665 & 0.04615 & 3.3e-05 & 27 & 10 & 9 & 968 \\
		\midrule
		2  & 1. ROUGE &  0.054714 & 0.987621 & \textbf{0.062322} & 1.3e-05 & 16 & 4 & 2 & 1033 \\
		130  & 2. ROUGE & 0.038864 & 0.99047 & 0.061189 & 3e-05 & 11 & 9 & 6 & 675 \\
		42  & 3. ROUGE & 0.04477 & 0.989132 & 0.05983 & 3.1e-05 & 12 & 11 & 2 & 852 \\
		\bottomrule
  \end{tabular}
  }
  \caption{List of the top three Trials of hyperparameter tuning for each evaluation metric (objective) along with the marking of the best values for each objective.}
  \label{tab:top3perEvalMetric}
\end{table}

\begin{table}[h]
    \scriptsize
    \centering
    \resizebox{\linewidth}{!}{
    \begin{tabular}{ccccccccccccc}
        \toprule
        \textbf{Trial Number} & \textbf{Loss$\downarrow$} & \textbf{Inverse Loss$\uparrow$ } & \textbf{BLEU$\uparrow$} & \textbf{ROUGE-$L$-$F1\uparrow$} & \textbf{Weighted Sum $f(x)$ $\uparrow$}\\
        \midrule
        116    &    0.0312    &    0.9688  &    0.9913    &    0.0461    &    \textbf{0.9801} \\
        127    &    0.0317    &    0.9683  &    0.9916    &    0.0465    &    0.9799 \\
        124    &    0.0324    &    0.9676  &    0.9917    &    0.0474    &    0.9797 \\
        108    &    0.0319    &    0.9681  &    0.9912    &    0.0488    &    0.9797 \\
        102    &    0.032    &    0.968   &    0.9913    &    0.0482     &    0.9797 \\
        \bottomrule
        \end{tabular}
    }
    \caption{List of the top five trials with the highest results for the weighted sum $f(x)$ sorted in descending order.}
    \label{tab:top5WeightedSum}
\end{table}

The maximum of the weighted sum $f(x)$ with $\theta^* = \max(f(x)) \mid x \in \text{Trials}$ is reached at trial 116. Concluding to $\theta^* = \theta_{f(\text{Trial}_{116})}$. Therefore, trial 116 defines the optimal hyperparameters \( \theta^* \) for the final training as follows: $\theta^*_{\text{learning\_rate}} = 3.4e-5 = 0.000035$, 
    $\theta^*_{\text{num\_train\_epochs}} = 6$,
    $\theta^*_{\text{warmup\_steps}} = 448$,
    $\theta^*_{\text{lora\_r\_value}} = 10$ and 
    $\theta^*_{\text{lora\_alpha\_value}} = 30$.

\subsubsection{Final Model Training:}

The final model training was conducted using the optimal hyperparameters, denoted as $\theta^*$, which were determined during the hyperparameter optimization phase. This training was performed on an NVIDIA RTX 2080 GPU with 11 GB of VRAM. Key statistics related to training times and memory usage are summarized in Table \ref{tab:finalTrainigResultsScores}. 

\begin{table}[h]
    \centering
    \resizebox{\linewidth}{!}{
    \begin{tabular}{ccccccccccccc}
        \toprule
        \textbf{Training Duration} & \textbf{Training Steps} & \textbf{Loss$\downarrow$ (Train Data)} & \textbf{Loss$\downarrow$ (Eval Data)} & \textbf{BLEU$\uparrow$ (Eval Data)} \\
        \midrule
        36.43hr   &    11.54k    &    0.0337 &    0.0393  &    0.9924    \\
        \bottomrule
        \toprule
        \textbf{Model Size (Quantized) } & \textbf{LoRA File Size} & \textbf{Train Dataset Size} & \textbf{Eval Dataset Size} & \textbf{Available VRAM}\\
        \midrule
        4.0046GB  &    21MB    & 30.3MB    &    7.4MB      &  11GB     \\
        \bottomrule
        \end{tabular}
    }
    \caption{Summary of the results of the final training.}
    \label{tab:finalTrainigResultsScores}
\end{table}

\begin{figure}[h]
	\centering
	\fbox{\includegraphics[width=0.8\linewidth]{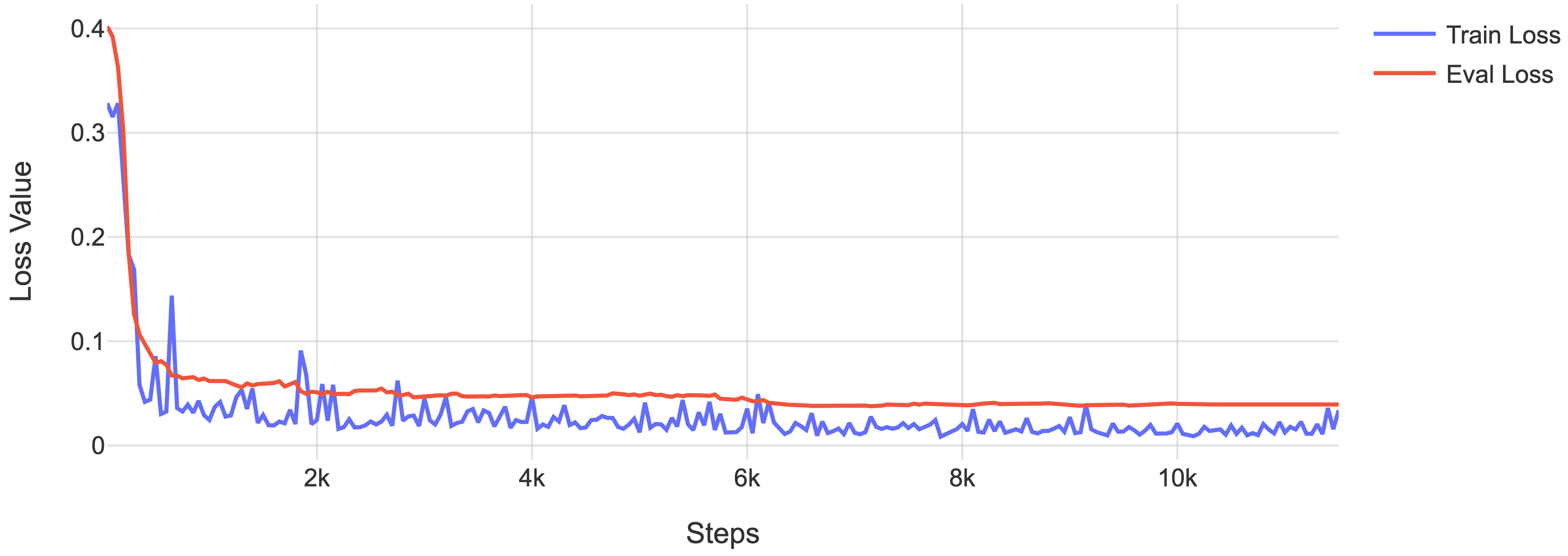}}
	\caption{The development of the Training Loss (blue) and Evaluation Loss (red) are plotted over the training steps of the final training, with an update every 50 steps.}
	\label{img:plot:loss}
\end{figure}

Figure \ref{img:plot:loss} illustrates the progression of the loss function for both the evaluation and test datasets. The training loss shows a high degree of fluctuation, while the evaluation loss remains stable throughout the process. Given that the evaluation loss follows a similar trend to the training loss without significant deviation, it can be inferred that overfitting did not occur during the training.

\subsubsection{Model Assessment}

In addition to the final training evaluation metrics (Table \ref{tab:finalTrainigResultsScores}), the results on the test dataset are shown in Table \ref{tab:EvalMetricsTestDataset}. A slight improvement in loss by 0.0028 and a minor decrease in BLEU by 0.0006 were observed, both considered negligible, indicating stable model performance.

During model assessment, hardware limitations restricted generation to 4,000 tokens per sample. Of 100 JSON samples from 20 prompts, only one terminated correctly within this limit. After post-processing to remove incomplete key-value pairs and close JSON objects, 81 out of 100 samples were successfully parsed.

Two prompt types were used: \textit{experimental} and \textit{clear} (see Section \ref{sec:ModelAssessmentMethods}). All 50 JSON samples from the \textit{clear} prompts were parsed without errors after post-processing. However, 19 parsing errors occurred in samples from the \textit{experimental} prompts.

The third phase involved qualitative analysis. Most samples were content-wise comparable to the original dataset, but limitations emerged. In some cases, the model repeated certain sections (e.g., Field Model) until reaching the 4k token limit. This repetitive behavior is permitted in JSON structure but unrealistic for real-world applications, requiring further investigation with more computational resources.

The analysis of the erroneously generated samples revealed two main types of issues:
\begin{enumerate}
    \item Some generated JSON objects began within another JSON object, such as within a key-value pair. This led to parsing problems due to violations of JSON syntax and structure. This error is likely caused by data chunking during preprocessing.
    \item The generation of unwanted characters, such as the zero-width space Unicode symbol (Unicode U+200B \cite{UnicodeStandardVersion}), was observed. Since these characters were not present in the training data, it is assumed that they are artifacts originating from the Code Llama 7B model.
\end{enumerate}

\begin{table}[ht!]
    \scriptsize
    \centering
    \resizebox{\linewidth}{!}{
    \begin{tabular}{ccccccccccccc}
        \toprule
        \textbf{Loss $\downarrow$} & \textbf{BLEU$\uparrow$} & \textbf{ROUGE-$1$-$F1\uparrow$} & \textbf{ROUGE-$2$-$F1\uparrow$} & \textbf{ROUGE-$L$-$F1\uparrow$} & \textbf{ROUGE-$L$-$F1\uparrow$} \\
        \midrule
        0.0309  &    0.9918    & 0.0565    &    0      &  0.0565   &  0.0565  \\
        \bottomrule
        \end{tabular}
    }
    \caption{Results of evaluation of the model from final training using the test dataset. In this case, the ROUGE values were not within the expected range and were included only for completeness.}
    \label{tab:EvalMetricsTestDataset}
\end{table}

\subsubsection{Results Conclusion:}
We can conclude that the model is capable of generating machine-readable JSON objects when given appropriate prompts. Fine-tuning on a GPU with sufficient memory produced strong results, and the abstraction of datasets reduced complexity, allowing for the use of customer data. This first version is already capable of being integrated into possible applications with respect to the described bias.

\section{Conclusions} 

The rapid development of AI technology and the increasing prevalence of \ac{LLMs} have created opportunities for new applications and tools. With continuous improvements in the efficiency of \ac{LLMs}, they are increasingly being used in the consumer and \ac{SME} sectors. This work demonstrates the potential and capabilities of open \ac{LLMs}, the extent to which the development of efficient model trainings with approaches such as \ac{LoRA} and quantization has already progressed and how these can be combined under restrictive resources. 

The work of this paper mark an initial step towards a generative AI assistance system showed the potential for further development into an assistant system for \ac{DDD} software development framework. In particular, the ability to generate machine-readable JSON objects enabled the use of the final model of this paper in potential tool chains and systems. The results and findings as well as the limitations and challenges, form a broad basis for further development.

Thus, the work of this thesis is a functional code generation model prototype that offers further possibilities and learning, paving the way for the development of an “\textit{Artificial Intelligent}” assistant system that meets the requirements and needs of \ac{DDD} software development.

The code we used to train and evaluate our models is available at \url{https://github.com/Tr33Bug/DomainlifecyclesCodeGenerator}. 


\newpage
\bibliographystyle{splncs}
\bibliography{sources}

@book{evansDomaindrivenDesignTackling2004,
  title = {Domain-Driven Design: Tackling Complexity in the Heart of Software},
  shorttitle = {Domain-Driven Design},
  author = {Evans, Eric},
  publisher = {Addison-Wesley},
  year = 2004
}

@book{evansDomainDrivenDesignReference2014,
  title = {Domain-{{Driven Design Reference}}: {{Definitions}} and {{Pattern Summaries}}},
  shorttitle = {Domain-{{Driven Design Reference}}},
  author = {Evans, Eric},
  publisher = {Dog Ear Publishing},
  year = 2014
}

@inproceedings{papineni-etal-2002-bleu,
    title = "{B}leu: a Method for Automatic Evaluation of Machine Translation",
    author = "Papineni, Kishore  and
      Roukos, Salim  and
      Ward, Todd  and
      Zhu, Wei-Jing",
    editor = "Isabelle, Pierre  and
      Charniak, Eugene  and
      Lin, Dekang",
    booktitle = "Proceedings of the 40th Annual Meeting of the Association for Computational Linguistics",
    month = jul,
    year = "2002",
    address = "Philadelphia, Pennsylvania, USA",
    publisher = "Association for Computational Linguistics",
    url = "https://aclanthology.org/P02-1040",
}

@misc{esentriInternalCompanyDocumentation,
  title = {Internal {{Company Documentation}}},
  author = {{esentri}, confidential},
  url = {https://esentri.com/},
  publisher = {esentri},
  year = 2024
}

@misc{huLoRALowRankAdaptation2021,
  title={LoRA: Low-Rank Adaptation of Large Language Models}, 
  author={Edward J. Hu and Yelong Shen and Phillip Wallis and Zeyuan Allen-Zhu and Yuanzhi Li and Shean Wang and Lu Wang and Weizhu Chen},
  year={2021},
  eprint={2106.09685},
  archivePrefix={arXiv},
  primaryClass={cs.CL},
  url={https://arxiv.org/abs/2106.09685}, 
}

@misc{dettmersQLoRAEfficientFinetuning2023,
      title={QLoRA: Efficient Finetuning of Quantized LLMs}, 
      author={Tim Dettmers and Artidoro Pagnoni and Ari Holtzman and Luke Zettlemoyer},
      year={2023},
      eprint={2305.14314},
      archivePrefix={arXiv},
      primaryClass={cs.LG},
      url={https://arxiv.org/abs/2305.14314}, 
}

@misc{HuggingFaceBitsAndBytes,
  author       = {{Hugging Face}},
  title        = {Bits and Bytes Documentation v0.44.1},
  year         = {2024},
  howpublished = {\url{https://huggingface.co/docs/bitsandbytes/main/en/index}},
  urldate = {2024-06-06},
}

@inproceedings{sunUltraLowPrecision4bit,
author = {Sun, Xiao and Wang, Naigang and Chen, Chia-yu and Ni, Jia-min and Agrawal, Ankur and Cui, Xiaodong and Venkataramani, Swagath and El Maghraoui, Kaoutar and Srinivasan, Vijayalakshmi and Gopalakrishnan, Kailash},
title = {Ultra-low precision 4-bit training of deep neural networks},
year = {2020},
isbn = {9781713829546},
publisher = {Curran Associates Inc.},
address = {Red Hook, NY, USA},
booktitle = {Proceedings of the 34th International Conference on Neural Information Processing Systems},
articleno = {152},
numpages = {12},
location = {Vancouver, BC, Canada},
series = {NIPS '20}
}

@misc{neshaeiImpactQuantizationRobustness2024,
  title={The Impact of Quantization on the Robustness of Transformer-based Text Classifiers}, 
  author={Seyed Parsa Neshaei and Yasaman Boreshban and Gholamreza Ghassem-Sani and Seyed Abolghasem Mirroshandel},
  year={2024},
  eprint={2403.05365},
  archivePrefix={arXiv},
  primaryClass={cs.CL},
  url={https://arxiv.org/abs/2403.05365}, 
}

@misc{roziereCodeLlamaOpen2024,
      title={Code Llama: Open Foundation Models for Code}, 
      author={Baptiste Rozière and Jonas Gehring and Fabian Gloeckle and Sten Sootla and Itai Gat and Xiaoqing Ellen Tan and Yossi Adi and Jingyu Liu and Romain Sauvestre and Tal Remez and Jérémy Rapin and Artyom Kozhevnikov and Ivan Evtimov and Joanna Bitton and Manish Bhatt and Cristian Canton Ferrer and Aaron Grattafiori and Wenhan Xiong and Alexandre Défossez and Jade Copet and Faisal Azhar and Hugo Touvron and Louis Martin and Nicolas Usunier and Thomas Scialom and Gabriel Synnaeve},
      year={2024},
      eprint={2308.12950},
      archivePrefix={arXiv},
      primaryClass={cs.CL},
      url={https://arxiv.org/abs/2308.12950}, 
}

@misc{liStarCoderMaySource2023,
      title={StarCoder: may the source be with you!}, 
      author={Raymond Li and Loubna Ben Allal and Yangtian Zi and Niklas Muennighoff and Denis Kocetkov and Chenghao Mou and Marc Marone and Christopher Akiki and Jia Li and Jenny Chim and Qian Liu and Evgenii Zheltonozhskii and Terry Yue Zhuo and Thomas Wang and Olivier Dehaene and Mishig Davaadorj and Joel Lamy-Poirier and João Monteiro and Oleh Shliazhko and Nicolas Gontier and Nicholas Meade and Armel Zebaze and Ming-Ho Yee and Logesh Kumar Umapathi and Jian Zhu and Benjamin Lipkin and Muhtasham Oblokulov and Zhiruo Wang and Rudra Murthy and Jason Stillerman and Siva Sankalp Patel and Dmitry Abulkhanov and Marco Zocca and Manan Dey and Zhihan Zhang and Nour Fahmy and Urvashi Bhattacharyya and Wenhao Yu and Swayam Singh and Sasha Luccioni and Paulo Villegas and Maxim Kunakov and Fedor Zhdanov and Manuel Romero and Tony Lee and Nadav Timor and Jennifer Ding and Claire Schlesinger and Hailey Schoelkopf and Jan Ebert and Tri Dao and Mayank Mishra and Alex Gu and Jennifer Robinson and Carolyn Jane Anderson and Brendan Dolan-Gavitt and Danish Contractor and Siva Reddy and Daniel Fried and Dzmitry Bahdanau and Yacine Jernite and Carlos Muñoz Ferrandis and Sean Hughes and Thomas Wolf and Arjun Guha and Leandro von Werra and Harm de Vries},
      year={2023},
      eprint={2305.06161},
      archivePrefix={arXiv},
      primaryClass={cs.CL},
      url={https://arxiv.org/abs/2305.06161}, 
}

@misc{wangCodeT5IdentifierawareUnified2021,
      title={CodeT5: Identifier-aware Unified Pre-trained Encoder-Decoder Models for Code Understanding and Generation}, 
      author={Yue Wang and Weishi Wang and Shafiq Joty and Steven C. H. Hoi},
      year={2021},
      eprint={2109.00859},
      archivePrefix={arXiv},
      primaryClass={cs.CL},
      url={https://arxiv.org/abs/2109.00859}, 
}

@misc{GitHubCopilotYour2024,
  author       = {{GitHub Inc.}},
  title        = {GitHub Copilot},
  year         = {2024},
  howpublished = {\url{https://github.com/features/copilot}},
  urldate = {2024-04-24},
}

@misc{KICodegeneratorAmazonCodeWhisperer,
  author       = {{Amazon Web Services (AWS)}},
  title        = {{AWS} CodeWhisperer},
  year         = {2024},
  howpublished = {\url{https://aws.amazon.com/de/codewhisperer/}},
  urldate = {2024-06-06},
}

@misc{louppeUnderstandingRandomForests2015,
      title={Understanding Random Forests: From Theory to Practice}, 
      author={Gilles Louppe},
      year={2015},
      eprint={1407.7502},
      archivePrefix={arXiv},
      primaryClass={stat.ML},
      url={https://arxiv.org/abs/1407.7502}, 
}

@inproceedings{tranDoesBLEUScore2019,
   title={Does BLEU Score Work for Code Migration?},
   volume={10},
   url={http://dx.doi.org/10.1109/ICPC.2019.00034},
   DOI={10.1109/icpc.2019.00034},
   booktitle={2019 IEEE/ACM 27th International Conference on Program Comprehension (ICPC)},
   publisher={IEEE},
   author={Tran, Ngoc and Tran, Hieu and Nguyen, Son and Nguyen, Hoan and Nguyen, Tien},
   year={2019},
   month=may, pages={165–176} }

@misc{chenEvaluatingLargeLanguage2021,
      title={Evaluating Large Language Models Trained on Code}, 
      author={Mark Chen and Jerry Tworek and Heewoo Jun and Qiming Yuan and Henrique Ponde de Oliveira Pinto and Jared Kaplan and Harri Edwards and Yuri Burda and Nicholas Joseph and Greg Brockman and Alex Ray and Raul Puri and Gretchen Krueger and Michael Petrov and Heidy Khlaaf and Girish Sastry and Pamela Mishkin and Brooke Chan and Scott Gray and Nick Ryder and Mikhail Pavlov and Alethea Power and Lukasz Kaiser and Mohammad Bavarian and Clemens Winter and Philippe Tillet and Felipe Petroski Such and Dave Cummings and Matthias Plappert and Fotios Chantzis and Elizabeth Barnes and Ariel Herbert-Voss and William Hebgen Guss and Alex Nichol and Alex Paino and Nikolas Tezak and Jie Tang and Igor Babuschkin and Suchir Balaji and Shantanu Jain and William Saunders and Christopher Hesse and Andrew N. Carr and Jan Leike and Josh Achiam and Vedant Misra and Evan Morikawa and Alec Radford and Matthew Knight and Miles Brundage and Mira Murati and Katie Mayer and Peter Welinder and Bob McGrew and Dario Amodei and Sam McCandlish and Ilya Sutskever and Wojciech Zaremba},
      year={2021},
      eprint={2107.03374},
      archivePrefix={arXiv},
      primaryClass={cs.LG},
      url={https://arxiv.org/abs/2107.03374}, 
}

@misc{yetistirenEvaluatingCodeQuality2023,
      title={Evaluating the Code Quality of AI-Assisted Code Generation Tools: An Empirical Study on GitHub Copilot, Amazon CodeWhisperer, and ChatGPT}, 
      author={Burak Yetiştiren and Işık Özsoy and Miray Ayerdem and Eray Tüzün},
      year={2023},
      eprint={2304.10778},
      archivePrefix={arXiv},
      primaryClass={cs.SE},
      url={https://arxiv.org/abs/2304.10778}, 
}

@misc{bourhisJSONDataModel2017,
      title={JSON: data model, query languages and schema specification}, 
      author={Pierre Bourhis and Juan L. Reutter and Fernando Suárez and Domagoj Vrgoč},
      year={2017},
      eprint={1701.02221},
      archivePrefix={arXiv},
      primaryClass={cs.DB},
      url={https://arxiv.org/abs/1701.02221}, 
}

@article{josephOptimalRatioData2022,
   title={Optimal ratio for data splitting},
   volume={15},
   ISSN={1932-1872},
   url={http://dx.doi.org/10.1002/sam.11583},
   DOI={10.1002/sam.11583},
   number={4},
   journal={Statistical Analysis and Data Mining: The ASA Data Science Journal},
   publisher={Wiley},
   author={Joseph, V. Roshan},
   year={2022},
   month=apr, pages={531–538} 
}

@misc{HuggingfaceTrainer,
  author       = {{Hugging Face}},
  title        = {Trainer Class for Fine-Tuning Models},
  year         = {2023},
  url          = {https://huggingface.co/docs/transformers/main_classes/trainer},
  urldate = {2024-04-29},
}

@inproceedings{papineniBleuMethodAutomatic2002,
    title = "{B}leu: a Method for Automatic Evaluation of Machine Translation",
    author = "Papineni, Kishore  and
      Roukos, Salim  and
      Ward, Todd  and
      Zhu, Wei-Jing",
    editor = "Isabelle, Pierre  and
      Charniak, Eugene  and
      Lin, Dekang",
    booktitle = "Proceedings of the 40th Annual Meeting of the Association for Computational Linguistics",
    month = jul,
    year = "2002",
    address = "Philadelphia, Pennsylvania, USA",
    publisher = "Association for Computational Linguistics",
    url = "https://aclanthology.org/P02-1040",
    doi = "10.3115/1073083.1073135",
    pages = "311--318",
}

@inproceedings{linROUGEPackageAutomatic2004,
    title = "{ROUGE}: A Package for Automatic Evaluation of Summaries",
    author = "Lin, Chin-Yew",
    booktitle = "Text Summarization Branches Out",
    month = jul,
    year = "2004",
    address = "Barcelona, Spain",
    publisher = "Association for Computational Linguistics",
    url = "https://aclanthology.org/W04-1013",
    pages = "74--81",
}

@misc{MethodsToolsEfficient,
  author       = {{Hugging Face}},
  title        = {Performance considerations for training on GPUs},
  year         = {2024},
  howpublished = {\url{https://huggingface.co/docs/transformers/main/en/perf_train_gpu_one}},
  urldate = {2024-04-25},
}

@misc{LoRAHuggingface,
  author       = {{Hugging Face}},
  title        = {LoRA (Low-Rank Adaptation) for Training Diffusion Models},
  year         = {n.d.},
  howpublished = {\url{https://huggingface.co/docs/diffusers/training/lora}},
  urldate = {2024-05-07},
}

@article{bazganPowerWeightedSum2022,
   title={The Power of the Weighted Sum Scalarization for Approximating Multiobjective Optimization Problems},
   volume={66},
   ISSN={1433-0490},
   url={http://dx.doi.org/10.1007/s00224-021-10066-5},
   DOI={10.1007/s00224-021-10066-5},
   number={1},
   journal={Theory of Computing Systems},
   publisher={Springer Science and Business Media LLC},
   author={Bazgan, Cristina and Ruzika, Stefan and Thielen, Clemens and Vanderpooten, Daniel},
   year={2021},
   month=nov, pages={395–415} }

@misc{UnicodeStandardVersion,
  author       = {{The Unicode Consortium}},
  title        = {The Unicode Standard, Version 15.0},
  year         = {2022},
  howpublished = {\url{https://www.unicode.org/versions/Unicode15.0.0/UnicodeStandard-15.0.pdf}},
  url = {https://www.unicode.org/versions/Unicode15.0.0/UnicodeStandard-15.0.pdf},
  urldate = {2024-06-06},
}



\end{document}